\documentclass{article}
\usepackage{spconf,amsmath,graphicx,hyperref}
\usepackage{svg}
\usepackage{booktabs, xcolor, makecell}
\usepackage{booktabs}
\usepackage{graphicx}
\usepackage{cite}
\usepackage[table]{xcolor}
\usepackage{float} 
\usepackage{adjustbox} 
\usepackage{indentfirst}
\usepackage{amsmath}
\usepackage{amssymb}
\usepackage{tabularx,array,xcolor,booktabs}
\newcolumntype{Y}{>{\centering\arraybackslash}X}

\usepackage{xcolor}

\title{TICL: Text-Embedding KNN For Speech In-Context Learning Unlocks Speech Recognition Abilities of Large Multimodal Models}
\name{Haolong Zheng, Yekaterina Yegorova, Mark Hasegawa-Johnson}
\address{ University of Illinois at Urbana-Champaign
\\\{haolong2, yay2, jhasegaw\}@illinois.edu}
\begin{document}
%
\maketitle
\begin{abstract}
Speech foundation models have recently demonstrated the ability to perform Speech In-Context Learning (SICL). 
Selecting effective in-context examples is crucial for SICL performance, yet selection methodologies remain under-explored. 
In this work, we propose Text-Embedding KNN for SICL (TICL), a simple pipeline that uses semantic context to enhance off-the-shelf large multimodal models' speech recognition abilities without fine-tuning. Across challenging automatic speech recognition tasks, including accented English, multilingual speech, and children’s speech, our method enable model to surpass zero-shot performance up to 84.7\% relative WER reduction. Ablation studies are conducted to show the robustness and efficiency of our method. 
\end{abstract}

\section{Introduction}
\label{sec:intro}


In-Context Learning (ICL) \cite{NEURIPS2020_1457c0d6} has been widely adopted for Large Language Models (LLMs), enabling adaptation to new tasks by conditioning on the demonstrations provided in the input context \cite{dong-etal-2024-survey}. It avoids costly fine-tuning which might lead to catastrophic forgetting issue.
A crucial factor that affects the ICL performance is the selection of in-context examples\cite{pmlr-v139-zhao21c, yang-etal-2023-representative, NEURIPS2024_8cb564df}.



Speech In-Context Learning (SICL) extends the idea to models that can process speech inputs, which was first introduced in \cite{10446502}.
It applied SICL to enhance Whisper's \cite{radford2022whisper} recognition of Chinese dialect, where k-nearest neighbors were retrieved as in-context examples using Whisper embeddings. They reduce the candidate dataset for each sample by searching for the most similar speaker with speaker embeddings. Their experiments show that the selection of in-context examples influences the performance of SICL. However, experiments were limited to word-level speech due to the computational complexity. To address this, \cite{zhou2025m2rwhispermultistagemultiscaleretrieval} proposed a sentence-level datastore, where utterance embeddings are mean-pooled and retrieved in a time-agnostic space. 
Nevertheless, Whisper’s relatively small context window restricted the number of examples that could be effectively incorporated.

    

\begin{figure}[!th]
    \centering
    \label{fig:pipeline}
    \includegraphics[width=\linewidth]{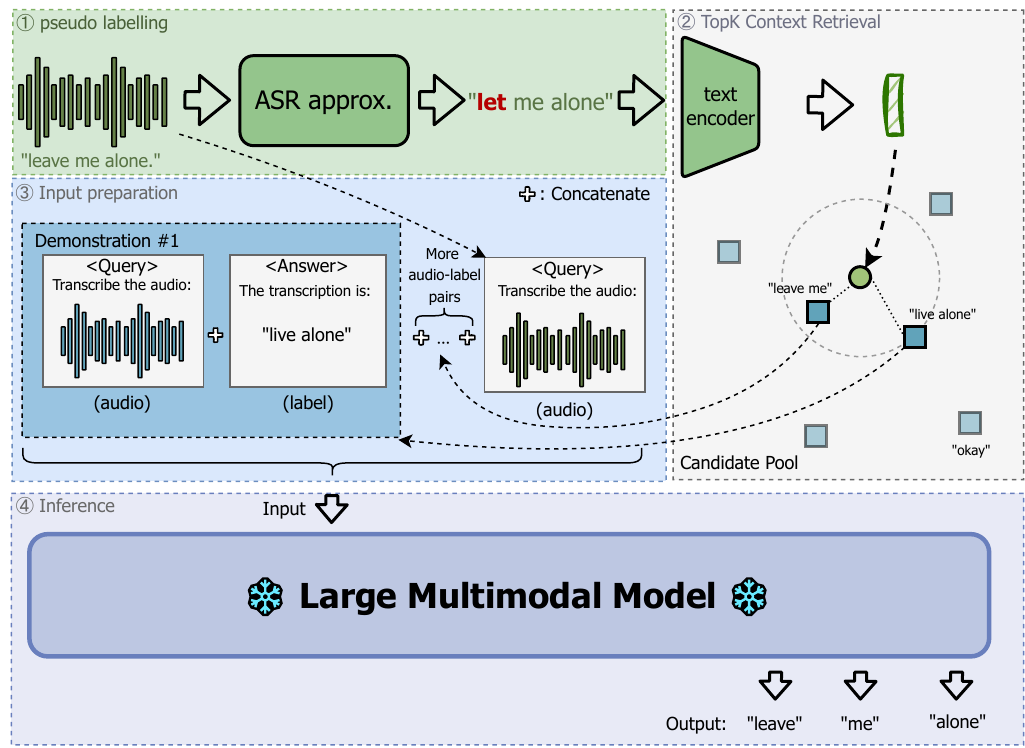}
    \caption{\textbf{Overview of the 
    TICL pipeline}:
  (1) A pre-trained ASR produces a pseudo-label for the test audio.
  (2) A text encoder embeds this pseudo-label and retrieves the 
  $K$ nearest candidates from a pool whose ground-truth transcripts are pre-embedded with the same encoder.
  (3) The retrieved 
  (audio, label) pairs are prepended as in-context learning examples prior to the test audio. 
  (4) Examples and test audio are passed as
  input to a large multimodal model (LMM) to generate the final 
  transcription output.
  }
\end{figure}



With the advent of large multimodal models (LMMs) that offer much larger context windows, SICL research has expanded. For example, \cite{roll2025incontextlearningboostsspeech} used Phi-4-Multimodal-instruct \cite{microsoft2025phi4minitechnicalreportcompact} and scaled the number of demonstrations up to 12. This yielded a relative 19.7\% performance improvement on accented English speech recognition. \cite{cheng2025speech} evaluated the SICL capability of Gemini-1.5 Pro\cite{team2024gemini} on accented speech with up to 100 demonstrations. However, both approaches relied on random sampling for SICL examples, which may underutilize the potential of SICL

To address the challenge of selecting the SICL examples, we introduce 
Text-Embedding KNN for SICL (TICL), a framework designed to leverage semantic retrieval of high-quality demonstrations for SICL, illustrated in Fig. \ref{fig:pipeline}.



\begin{table}[t]
\centering
\caption{
TICL Accented English Results. \(\downarrow\)WER\%}
\label{tab:accent-en-wer}
\resizebox{0.95\linewidth}{!}{%
\begin{tabular}{@{}lcccc@{}}
\toprule
& \multicolumn{2}{c}{\textbf{GLOBE-V2}} & \multicolumn{2}{c}{\textbf{L2-Arctic}} \\
\cmidrule(lr){2-3}\cmidrule(lr){4-5}
& \textbf{\textsc{Phi-4-MM}} & \textbf{Qwen2-Audio} & \textbf{\textsc{Phi-4-MM}} & \textbf{Qwen2-Audio} \\
\midrule
\rowcolor{white}
$\mathbf{k{=}0}$ & 4.23 & 5.41 & 8.47 & 11.06 \\
\cmidrule[\heavyrulewidth]{1-5}
\rowcolor{gray!10}
$k{=}1$ & 1.40 & 2.38 & 2.94 & 4.76 \\
\rowcolor{white}
$k{=}2$ & 1.13 & 2.67 & 2.81 & 1.7 \\
\rowcolor{gray!10}
$k{=}3$ & 0.92 & 1.89 & 2.70 & 1.52 \\
\rowcolor{white}
$k{=}4$ & \textbf{0.88} & \textbf{1.66} & \textbf{2.62} & \textbf{1.41} \\
\cmidrule[\heavyrulewidth]{1-5}
\rowcolor{white}
\emph{$\Delta_{\mathrm{rel}}$} & \textbf{79.2\%} & \textbf{69.3\%} & \textbf{69.1\%} & \textbf{84.7\%} \\
\bottomrule
\end{tabular}%
}
\end{table}

\section{Methodology}
\label{sec:methodology}

\subsection{Speech In-Context Learning}
\label{sec:SICL}
ICL can adapt the model through conditioning on a given demonstration comprising of target domain data instead of updating the model weights. Unlike purely textual ICL, SICL conditions on audio and text tokens at the same time. For SICL adapted ASR, given a test speech sample $\boldsymbol{s^*}$, a model $\Lambda$ generates the predicted transcription $\boldsymbol{\hat{y}}$ as: 
\[ \boldsymbol{\hat{y}} = \arg\max_{\boldsymbol{y}} \mathrm{Pr}(\boldsymbol{y} \mid C, \boldsymbol{x_s^*}, \Lambda),\]
where $\boldsymbol{x_s^*}$ corresponds to the audio encoding of $\boldsymbol{s^*}$ and $C$ denotes the context 
Generally, $C$ is structured as a dialogue history\footnote{Preliminary experiments show that severe hallucinations will occur if we model $\boldsymbol{\hat{y}} = \arg\max_{\boldsymbol{y}} P_{\text{SICL}}(\boldsymbol{y} \mid \boldsymbol{X_a}, \boldsymbol{Y_c}, \Lambda)$ where $\boldsymbol{X_a}=[\boldsymbol{x_a^{(1)};\dots;x_a^{(i)};x_a^*}]$ and $\boldsymbol{Y_c}=[\boldsymbol{y^{(1)};\dots;y{(i)}}]$.} of query–answer pairs $c^{(i)}=(q^{(i)}, a^{(i)})$. In our case, query $q$ is the concatenation of an encoded text prompt $\boldsymbol{x_p^{(i)}}$ and an encoded audio segment 
$\boldsymbol{x_s^{(i)}}$, 
while the answer ${a}$ is the transcription of $\boldsymbol{s^{(i)}}$, denoted by
$\boldsymbol{y^{(i)}}$.
Therefore, with slight abuse of notation, the full context $C$ in our case can be denoted as:

$
    C = [c^{(1)};c^{(2)};\dots;c^{(n)}], \hspace{12pt} c^{(i)}=[\boldsymbol{x_p^{(i)}}; \boldsymbol{x_s^{(i)}};\boldsymbol{y^{(i)}}]. 
$








\subsection{Text-Embedding 
KNN Candidate Selection}
The goal of this work is to find an appropriate context $C$ for a given test sample. \cite{sia-duh-2023-context, zebaze2025context} show that models benefit the most from the context when ICL examples are from the same domain as the test sample. Motivated by this, we choose SICL examples by selecting the sample with transcription $\boldsymbol{y^{(i)}}$ that is lexically similar to the test audio's transcription $\boldsymbol{y^{*}}$.
To obtain the closest samples in the lexicon space, we use a pre-trained text encoder to encode the transcription and retrieve the $K$ nearest candidates using the Euclidean distance. However, during inference, $\boldsymbol{y^{*}}$ is not available. Therefore, we use a pre-trained ASR model to generate a pseudo-label $\boldsymbol{\tilde{y}}$, which can include errors, and then we encode the pseudo-label to get an approximate embedding.  

More specifically: 
Let the candidate dataset be
$\mathcal{C}=\{(\boldsymbol{s^{(i)}}, \boldsymbol{y^{(i)}})\}_{i=1}^N$, where $\boldsymbol{s^{(i)}}$ and ${\boldsymbol{y^{(i)}}}$ is the speech audio and its transcription respectively. 
Let $\phi:\mathcal{Y}_{\text{text}}\to\mathbb{R}^d$ be a frozen sentence embedding model which maps any text sentence from the set of all sentences $\mathcal{Y}_\mathrm{text}$ to a $d-$dimensional vector. We define $\bar{\phi}(\boldsymbol{y})=\phi(\boldsymbol{y})/\|\phi(\boldsymbol{y})\|_2$ as the $\ell_2$-normalized embedding. For each candidate $c\in\mathcal{C}$ we precompute $\bar{\boldsymbol{z}}=\bar{\phi}(\boldsymbol{y})$.
Let $f_\theta:\mathcal{X}_{\text{audio}}\to\mathcal{Y}_{\text{text}}$ denote a frozen ASR model (e.g., Whisper). 
During inference, given an audio $\boldsymbol{s^\ast}$, we first obtain a pseudo-transcription
$\boldsymbol{\tilde{y}} = f_\theta(\boldsymbol{s^\ast})$
and compute its lexical embedding $\boldsymbol{\bar{z}^\ast}=\bar{\phi}(\boldsymbol{\tilde{y}})$.  We select the $K$ nearest candidates with Euclidean distance in the normalized embedding space,
\[
\mathcal{N}_K(\boldsymbol{s^\ast})\;=\;\operatorname{TopK}_{i\in[N]}\!\bigl(-r(i)\bigr),
\]
where
$r(i) \;=\; \bigl\|\boldsymbol{\bar{z}^\ast} - \boldsymbol{\bar{z}^{(i)}}\bigr\|_2.$
Finally,  using the $K$ candidates we construct the context $C$ according to \ref{sec:SICL}.

\begin{figure}[t]
    \centering
    \caption{Retrieval Method Comparison Result}
    \includegraphics[width=0.78\linewidth]{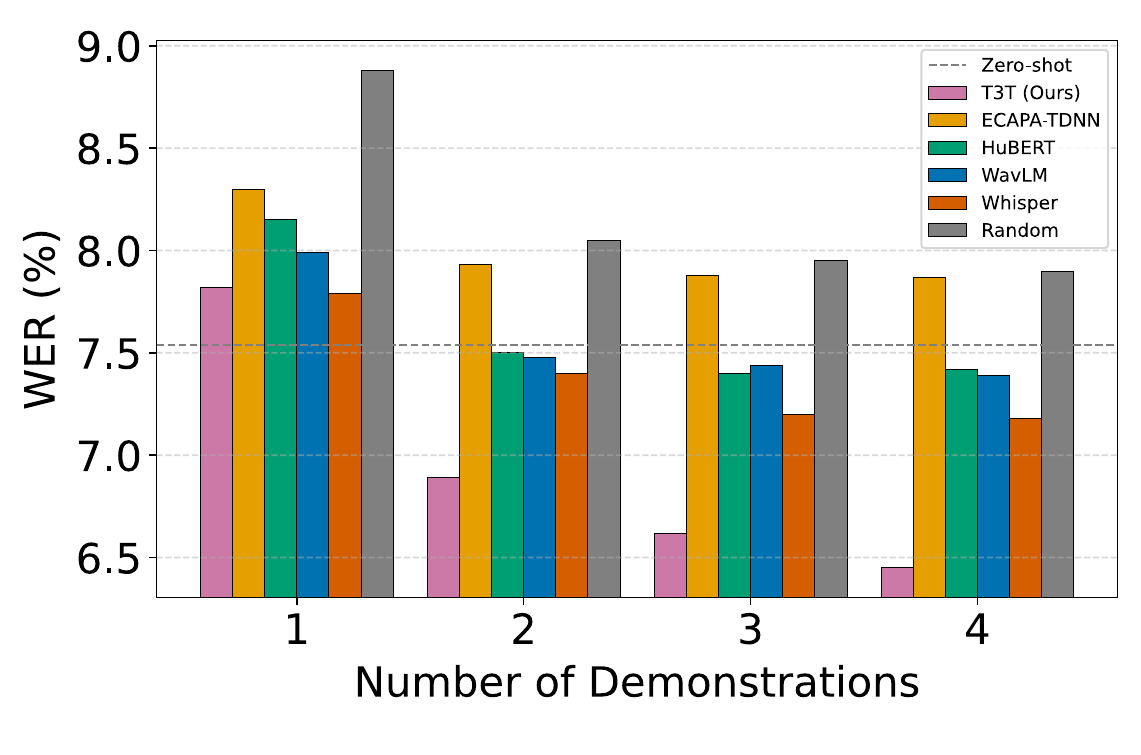}
    \label{fig:baseline}
\end{figure}
\section{Experiments}
\label{sec:Experiment}

\begin{table*}[!ht]
\centering
\ninept
\caption{TICL Multilingual Results for \emph{Phi-4-MM}. \(\downarrow\)WER\% by default and \(\downarrow\)CER\% for zh/ja/th.}
\label{tab:multilingual}
\resizebox{\linewidth}{!}{%
\begin{tabular}{@{}lccccccccccccc@{}}
\toprule
& \multicolumn{8}{c}{\textbf{Natively Supported Languages}} & \multicolumn{5}{c}{\textbf{Unsupported Languages}} \\
\cmidrule(lr){2-9}\cmidrule(lr){10-14}
& \textbf{de} & \textbf{en} & \textbf{es} & \textbf{fr} & \textbf{it} & \textbf{ja} & \textbf{pt} & \textbf{zh}
& \textbf{nl} & \textbf{pl} & \textbf{ru} & \textbf{th} & \textbf{tr} \\
\midrule
\rowcolor{white}
$\mathbf{k{=}0}$ & \textbf{5.24} & 7.56 & \textbf{4.27} & 8.00 & 3.79 & 13.00 & 6.06 & \textbf{8.49} & 101.15 & 117.55 & 122.75 & 134.21 & 132.74 \\
\cmidrule[\heavyrulewidth]{1-14}
\rowcolor{gray!10}
$k{=}1$ & 8.82 & 7.82 & 6.66 & 8.08 & 4.79 & 8.11 & 3.60 & 13.15 & 58.41 & 37.91 & 21.49 & 68.23 & 39.87 \\
\rowcolor{white}
$k{=}2$ & 6.16 & 6.89 & 5.78 & 7.63 & 4.09 & 7.06 & 3.73 & 12.67 & \textbf{58.37} & \textbf{35.58} & 19.82 & \textbf{64.51} & \textbf{36.34} \\
\rowcolor{gray!10}
$k{=}3$ & 5.75 & 6.62 & 5.74 & 7.48 & 3.75 & 6.38 & 3.63 & 11.86 & 60.78 & 35.66 & \textbf{18.86} & 65.42 & 36.47 \\
\rowcolor{white}
$k{=}4$ & 5.45 & \textbf{6.45} & 5.63 & \textbf{7.41} & \textbf{3.64} & \textbf{6.17} & \textbf{3.52} & 11.07 & 63.10 & 37.22 & 20.74 & 65.78 & 37.15 \\
\cmidrule[\heavyrulewidth]{1-14}
\rowcolor{white}
\emph{$\Delta_{\mathrm{rel}}$}
& -4.0\% & \textbf{14.7\%} & -31.9\% & \textbf{7.4\%} & \textbf{4.0\%} & \textbf{52.5\%} & \textbf{41.9\%} & -30.4\%
& \textbf{42.3\%} & \textbf{69.7\%} & \textbf{84.6\%} & \textbf{51.9\%} & \textbf{72.6\%} \\
\bottomrule
\end{tabular}%
}
\end{table*}

The proposed pipeline is designed to be applicable across speech recognition tasks and compatible with any LMM. In this work, we apply it to Phi-4-MultiModal-instruct (Phi-4-MM) \cite{microsoft2025phi4minitechnicalreportcompact} and, in part, to Qwen2-Audio-7B-Instruct (Qwen2-Audio)\cite{Qwen2-Audio}. Unless otherwise noted, we use \textit{Whisper-Large-v3-turbo} as the pseudo-labeling ASR, due to its accuracy and quick runtime. For lexical retrieval, we represent each transcript with a sentence embedding using \texttt{all-mpnet-base-v2}.\footnote{\href{https://huggingface.co/sentence-transformers/all-mpnet-base-v2}{\texttt{all-mpnet-base-v2}}} 
Since semantically similar sentences are near each other in the embedding space, it recovers phrase- and intent-matched examples despite small pseudo-label imperfections.
In multilingual settings, we switch to \texttt{paraphrase-\allowbreak multilingual-\allowbreak mpnet-\allowbreak base-\allowbreak v2\footnote{\href{https://huggingface.co/sentence-transformers/paraphrase-multilingual-mpnet-base-v2}{ \texttt{paraphrase-multilingual-mpnet-base-v2}}}} to preserve these properties in a language-agnostic space. We evaluate using Word Error Rate (WER)
by default unless specified. 
 Our experiments include datasets\footnote{Experiments are restricted to utterances with durations between 1–15 seconds.} of accented English, multilingual speech, and children’s speech. All results are reported on the official test splits.

\subsection{Retrieval Method Comparison}


We systematically compare different retrieval strategies on the English subset of Common Voice. This dataset was chosen for its breadth of accents, topics, speaking styles, and recording conditions, which provide a large and diverse retrieval pool. In addition to two baselines, a Whisper-embedding retrieval method \cite{zhou2025m2rwhispermultistagemultiscaleretrieval} and uniform random selection of in-context examples \cite{roll2025incontextlearningboostsspeech}, we also evaluate embeddings that capture different dimensions of similarity. Specifically, we include HuBERT \cite{hsu2021hubertselfsupervisedspeechrepresentation} (content-oriented, approximating phonetic/lexical similarity), ECAPA-TDNN \cite{Desplanques_2020} (speaker-identity embeddings that are relatively content-invariant), and WavLM \cite{Chen_2022} (target-speaker embeddings that integrate identity and content cues while suppressing environmental noise and interfering speech). For HuBERT and WavLM, we use the final-layer hidden state as the utterance embedding. For embeddings with a termporal dimension, we apply statistical pooling over time to produce a fixed-length vector.


Fig.~\ref{fig:baseline} shows that TICL consistently achieves the best performance across all settings. Content- and phonetic-based retrieval methods (Whisper, HuBERT, WavLM) outperform speaker-based retrieval (ECAPA-TDNN), while the random baseline performs worst. The results again confirm that the quality of the context can influence the performance of SICL.

\subsection{Accented English Speech Recognition}
For accented English speech recognition we use GLOBE-V2 \cite{wang2024globe}, which contains 164 distinct accents, and L2-ARCTIC \cite{zhao18b_interspeech}, which contains read speech from non-native English speakers spanning six L1 backgrounds. In-context examples are retrieved from the training and validation splits. As demonstrated in Table \ref{tab:accent-en-wer}, our TICL retrieval method shows a generalizable and consistent improvement, up to 84.7\% compared to zero-shot.

\subsection{Multilingual Speech Recognition}
For multilingual speech recognition, we used the Common Voice Corpus 15.0 \cite{commonvoice:2020} dataset. To evaluate the impact to Phi-4-MM’s native ASR capabilities, we focused on the directly supported languages: English (en), Chinese (zh), German (de), French (fr), Italian (it), Japanese (ja), Spanish (es), and Portuguese (pt). In addition, Russian (ru), Dutch (nl), Polish (pl), Thai (th), and Turkish (tr) were included, which Phi-4-MM’s ASR does not support natively. 
In-context examples are retrieved from the validated split.

From Table~\ref{tab:multilingual}, 
our TICL pipeline is effective for the supported languages.
We see noticeable improvement for ja
and pt 
However, de, es and zh performance worsened. Through manually revisiting the results of zh, we found that many retrieved in-context examples were not lexically close to the target transcripts. Two factors contribute: (i) the dataset contains compound/rare terms, for which the pseudo-labeler often produces approximate transcripts with important mistakes, degrading the lexical query; and (ii) sentence-level multilingual-MPNet embeddings under-represent character/subword overlap for such items, causing the retriever to miss close matches and drift to unrelated examples. 

Table~\ref{tab:multilingual} further demonstrates that our TICL pipeline enables the model to transcribe languages that was not originally supported. Significant improvements are observed across all unsupported languages, with strong gains for ru, tr, and pl. These results suggest, with proper setup, SICL can unlock models' abilities on unseen task

\subsection{Children's Speech Recognition}
To evaluate children's speech recognition performance, we used four corpora: My Science Tutor (MyST) \cite{pradhan2024my}, the OGI Kids' Speech Corpus \cite{ogi}, Edmonton Narrative Norms Instrument (ENNI) \cite{schneider2006storytelling} and the Redmond Sentence Recall (RSR) \cite{ai4exceptionaled_rsr_hf}. We preprocess MyST and OGI following \cite{fan2024benchmarking} and ENNI following \cite{liu2024fasaflexibleautomaticspeech}. In-context examples are retrieved from the training and validation splits. As shown in Table~\ref{tab:child-wer}, TICL provides consistent improvements over the zero-shot baseline across all datasets. The largest improvement is on OGI, likely due to a better-matched candidate pool. For MyST, ENNI, and RSR the gains are more modest, 
which we attribute to the smaller or less diverse candidate pools.

\begin{table}[!t]
\centering
\ninept
\caption{TICL Children's Speech Results for \emph{Phi-4-MM}. \(\downarrow\)WER\%.}
\label{tab:child-wer}
\begin{tabularx}{\linewidth}{@{}lYYYY@{}}
\toprule
& \textbf{MyST} & \textbf{OGI} & \textbf{ENNI} & \textbf{RSR} \\
\midrule
\rowcolor{white}
$\mathbf{k{=}0}$ & 12.81 & 16.17 & 14.37 & 20.06 \\
\cmidrule[\heavyrulewidth]{1-5}
\rowcolor{gray!10}
$k{=}1$ & 17.27 & 9.55 & 17.57 & 18.92 \\
\rowcolor{white}
$k{=}2$ & 11.77 & 8.94 & 14.07 & 18.92 \\
\rowcolor{gray!10}
$k{=}3$ & \textbf{11.69} & 8.75 & \textbf{13.54} & \textbf{18.90} \\
\rowcolor{white}
$k{=}4$ & 11.81 & \textbf{8.52} & 13.75 & 19.54 \\
\cmidrule[\heavyrulewidth]{1-5}
\rowcolor{white}
\emph{$\Delta_{\mathrm{rel}}$} & \textbf{8.7\%} & \textbf{47.3\%} & \textbf{5.8\%} & \textbf{5.8\%} \\
\bottomrule
\end{tabularx}
\end{table}

\section{Ablation study}
\label{sec:ablation}

We conduct two ablation studies to analyze the effect of pseudo-labeler accuracy and the number of in-context examples on our proposed pipeline. We use the GLOBE-V2 dataset, which provides a good balance between size and quality. To ensure that our findings generalize across different LMM architectures, we evaluate with both Phi-4-MM and Qwen2-Audio.

\subsection{Sensitivity of the Pseudo-Labeler}

\begin{table}[!b]
\centering
\label{tab:pseudolabeler}
\caption{Impact of pseudo-labeler quality on TICL (\(K{=}4\)) for \textsc{GLOBE-V2}. $\downarrow$ WER\%. Parentheses show relative WER reduction vs.\ zero-shot (\(k{=}0\)) for the same model. 
}
\label{tab:pseudo-labeler-sensitivity}
\resizebox{\linewidth}{!}{%
\renewcommand{\arraystretch}{0.95}
\begin{tabular}{@{}lccc@{}}
\toprule
\textbf{Whisper Conf.} & \textbf{Pseudo-label WER} & \textbf{Phi-4MM} & \textbf{Qwen2-Audio-7B} \\
\midrule
\rowcolor{white}
$k{=}0$ & -- & 4.23 & 5.41 \\
\cmidrule[\heavyrulewidth]{1-4}
\rowcolor{gray!10}
tiny            & 13.11 & 1.36 {\ninept(\(\downarrow\) 67.85\%)} & 2.37 {\ninept(\(\downarrow\) 56.19\%)} \\
\rowcolor{white}
base            &  8.83 & 1.22 {\ninept(\(\downarrow\) 71.16\%)} & 2.14 {\ninept(\(\downarrow\) 60.44\%)} \\
\rowcolor{gray!10}
small           &  4.64 & 1.10 {\ninept(\(\downarrow\) 74.00\%)} & 1.91 {\ninept(\(\downarrow\) 64.70\%)} \\
\rowcolor{white}
medium          &  2.86 & 1.01 {\ninept(\(\downarrow\) 76.12\%)} & 1.88 {\ninept(\(\downarrow\) 65.25\%)} \\
\rowcolor{gray!10}
large           &  2.17 & 0.92 {\ninept(\(\downarrow\) 78.25\%)} & 1.82 {\ninept(\(\downarrow\) 66.36\%)} \\
\rowcolor{white}
large-v3-turbo  &  \textbf{1.95} & \textbf{0.88} {\ninept(\(\downarrow\) \textbf{79.20}\%)} & \textbf{1.66} {\ninept(\(\downarrow\) \textbf{69.32}\%)}\\
\bottomrule
\end{tabular}%
}
\end{table}

To evaluate how pseudo-label quality affects the TICL pipeline we simulate a range of pseudo-label qualities. Pseudo-labels are generated using Whisper configurations of various sizes (tiny, base, small, medium, large, large-v3-turbo), yielding pseudo-label WERs from 13.11\% down to 1.95\%. For each setting, TICL retrieves \(K=4\) in-context examples, and run on all selected LMMs. As shown in Fig.~\ref{tab:pseudolabeler}, higher quality pseudo-labels consistently give better performance. Notably, even with the noisiest labels,
TICL still surpasses the zero-shot baselines by a wide margin, while the overall performance gap between the best and worst pseudo-labelers configuration remains modest.
This suggests that TICL's sensitivity to the pseudo labeler is low. We attribute this robustness to lexical retrieval in embedding space: near-synonymous strings (e.g., “let me alone’’ vs. “leave me alone’’) remain close, enabling retrieval to mitigate transcription noise. Thus, although the pseudo-labeler introduces a potential bottleneck, its influence on TICL is limited.



\subsection{Number of In-Context Examples}
\begin{figure}[!t]
    \centering
    \caption{Impact of the number of demonstrations in the T3T SICL pipeline. Evaluated on GLOBE-V2 dataset.}
    \includegraphics[width=0.8\linewidth]{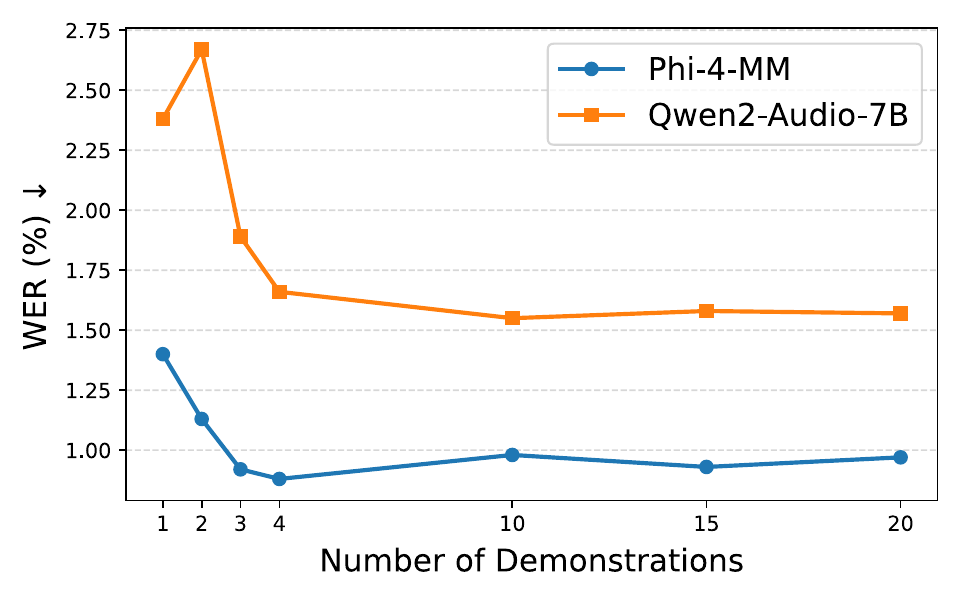}
    
    \label{fig:kshotwer}
\end{figure}

To investigate whether more demonstrations help, we also evaluated our pipeline with $K\in\{1,2,3,4,10,15,20\}$. As shown in Fig.~\ref{fig:kshotwer}, increasing $K$ beyond 
$K\approx 4$
adds little benefit and can hurt accuracy. We believe two effects are at play: (i) 
TICL
already efficiently surfaces the most useful, semantically close examples, so later additions contribute more noise than signal; and (ii) audio has a higher frame rate than text input, and long prompts degrade LMMs' performance because these models are predominantly trained on shorter sequences. In practice, the TICL pipeline works well with just a few demonstrations.

\section{Conclusion}
\label{sec:conclusion}
In this work, we proposed an enhanced and generalizable SICL pipeline called TICL. Our approach yields substantial performance gains, up to 84.7\% relative improvement, across three different speech recognition tasks. We conduct ablation studies showing the robustness and efficiency of our proposed retrieval method. This work offers a lightweight and cost-effective alternative for adapting models to new domains, especially for ASR. However, we observed limitations when datasets contained rare terms or complex vocabulary. Future work will explore strategies to address these failure cases, and study the mechanisms of the SICL paradigm.

\section{Acknowledgments}
This work was supported by National Science Foundation grant \#2229873. 
This work used the Delta system at the National Center for Supercomputing Applications through allocation beiq-delta-gpu from the Advanced Cyberinfrastructure Coordination Ecosystem: Services \& Support (ACCESS) program, which is supported by National Science Foundation grants \#2138259, \#2138286, \#2138307, \#2137603, and \#2138296.

{\ninept
\bibliographystyle{IEEEbib}
\bibliography{strings}
}

\end{document}